\documentclass[prb,twocolumn,showkeys,showpacs,preprintnumbers,amsmath,amssymb]{revtex4}
\usepackage{graphicx}
\usepackage{dcolumn}
\usepackage{bm}
\usepackage[tight]{subfigure}
\usepackage{enumerate}
\usepackage{amsmath}
\usepackage{verbatim}
\usepackage{color}
\usepackage{natbib}

\begin{document}
\title{Single-electron transport in InAs nanowire quantum dots formed by crystal phase engineering }
\today
\author{Malin Nilsson$^{1}$}
\author{Luna Namazi$^{1}$}
\author{Sebastian Lehmann$^{1}$}
\author{Martin Leijnse$^{1}$}
\author{Kimberly A. Dick$^{1,2}$}
\author{Claes Thelander$^{1}$}

\affiliation{$^{1}$Division of Solid State Physics and NanoLund, Lund University, Box 118,S-221 00 Lund, Sweden}
\affiliation{$^{2}$Center for Analysis and Synthesis, Lund University, Box 124, S-221 00 Lund, Sweden} 
  
  
\begin{abstract}
We report electrical characterization of quantum dots  formed by introducing pairs of thin wurtzite (WZ) segments in zinc blende (ZB) InAs nanowires. Regular Coulomb oscillations are observed over a wide gate voltage span, indicating that WZ segments create significant barriers for electron transport. We find a direct correlation of transport properties with quantum dot length and corresponding growth time of the enclosed ZB segment. The correlation is made possible by using a  method to extract lengths of nanowire crystal phase segments directly from scanning electron microscopy  images, and with support from transmission electron microscope images of typical nanowires. From experiments on controlled filling of nearly empty dots with electrons, up to the point where Coulomb oscillations can no longer be resolved, we estimate a lower bound  for the ZB-WZ conduction-band offset of  95~meV.
\end{abstract}

\maketitle

\section{Introduction}
Epitaxially grown semiconductor nanowires often have an uncontrolled mix of crystal phases, most commonly wurtzite (WZ) and zinc blende (ZB). Considerable efforts have been made to find growth conditions where a single, defect-free crystal phase is formed, which is required in many applications in electronics and optics. Of special interest is the WZ phase, which is unique to nanowires for the non-nitride III-V semiconductors. At the same time, new methods for crystal-phase control in nanowires have been used to controllably switch crystal phase in the growth direction of a single nanowire, creating a so-called homostructure.\cite{Caroff2009}  Importantly, such crystal phase junctions can be tuned to be atomically abrupt, and can be formed e.g. by metal-organic vapor phase epitaxy (MOVPE).

Homostructures consisting of thin segments of ZB and WZ have been optically studied in InP and GaAs nanowires,\cite{Akopian2010,Vainorius2015} resulting in a further understanding of effects of polytypism on the band structure in those material systems. In the case of InAs, a material with strong relevance for future low-power electronics, it has been theoretically predicted that WZ has a larger bandgap than ZB,\cite{Zanolli2007} with a positive conduction-band offset of up to 126~meV.\cite{Murayama1994,Belabbes2012} Mixing of the crystal phases in a nanowire device is therefore expected to strongly affect the electrical properties.\cite{Schroer2010,Thelander2011} 

Efforts have been made to study the band alignment at the interface between ZB and WZ in InAs nanowires, yet few experimental results have been reported to support the theoretical predictions, partly because of the difficulties involved in optical studies due to the narrow band gap ($E_{g} = 0.354$~eV in ZB at 295~K). Another complication is that the polar $\lbrace0001\rbrace/\lbrace000\bar{1}\rbrace$-type surfaces of WZ are expected to have polarization charges at the interfaces to the ZB, which should modify the band structure at the junction.\cite{Dayeh2009,Belabbes2013} However, for InAs, the amount of polarization charge relative to surface charges is unknown. Recent results from surface analysis of hydrogen-cleaned InAs nanowires show no detectable polarization charges at the interface or offset in the conduction band; the latter was explained by the intrinsic $n$-type characteristic of  InAs masking the fundamental offset.\cite{Hjort2014} In addition, the authors in Ref.~\onlinecite{Hjort2014} point out the importance of surface states and surface oxides for the band alignment. In experiments on InAs nanowires \textit{with} native oxides, thermionic emission measurements indicate that WZ forms barriers relative to ZB for electron transport.\cite{Thelander2011} The latter is also supported by the observation that very thin WZ segments in an otherwise ZB InAs nanowire act as tunnel barriers for electrons; pairs of such WZ barriers enclosing a ZB segment show single-electron charging at low temperatures, thus forming a quantum dot (QD).\cite{Dick2010} 

The electronic properties of InAs WZ-ZB junctions have become even more relevant by recent demonstrations that radial heterostructures can be formed selectively on ZB segments, whereas radial growth is effectively suppressed on WZ.\cite{Kawaguchi2011,Rieger2012,Namazi2015} Nanowire core templates consisting of InAs double-barrier WZ structures could thus be used to realize enclosed ZB quantum dots with an InAs-GaSb core-shell geometry, which would be of considerable interest for studies of electron-hole interactions.\cite{Ganjipour2015}

In this work, we present results on InAs nanowire homostructure QDs using a growth method that, compared to Ref.~\onlinecite{Dick2010}, involves higher growth temperature and an improved method to induce structural changes by change of group V/III precursor flow ratio.\cite{Lehmann2013,Namazi2015}  The new conditions allow a reduction in both impurity incorporation and QD size, which facilitates observation of Coulomb blockade and various quantum confinement-related effects. By correlating the length of ZB QDs with the frequency of Coulomb oscillations induced by the gate, we can confirm the role of the WZ segments as tunnel barriers. Coulomb blockade is observed over a wide range of gate voltages which, converted to an energy scale, provides a lower estimate of the conduction-band offset between WZ and ZB of around +95~meV.

Furthermore, as transmission electron microscopy (TEM) characterization is extremely difficult to carry out on nanowires on which three-terminal electrical measurements are performed, we have developed an approach to accurately extract lengths and volumes of nanowire crystal phase segments using electron channeling contrast imaging (ECCI) in scanning electron microscopy (SEM).\cite{Joy1982}

\section{Method}
InAs nanowires containing crystal-phase-defined QDs were grown by means of MOVPE on ($\bar{1} \bar{1} \bar{1}$)-oriented InAs substrates with predeposited aerosol Au seed particles. The particles used for this study were 40~nm in diameter with an areal density of 1 particle per~$\mu$m$^{2}$. A low-pressure MOVPE reactor from Aixtron with a total precursor flow of 13~l/min and operated at a pressure of 100~mbar was utilized with trimethylindium (TMIn) and arsine (AsH$_{3}$) as precursors. Prior to growth, the samples were annealed under H$_{2}$/AsH$_{3}$ at an elevated set temperature of 550$^{\circ}$C to remove native oxides from the surface of the substrate. Subsequently, the growth temperature was set to 460$^{\circ}$C. Different V/III ratios were employed for growing the WZ and ZB crystal structures by changing the molar fraction of both growth species; AsH$_{3}$ was set to $9.23\times10^{-5}$, and $1.54\times10^{-2}$, while TMIn was set to $3.48\times10^{-6}$, and $1.93\times10^{-6}$ for the WZ and ZB segments, respectively. The lengths of the segments were scaled and controlled with the set growth time. The two segments of WZ were both grown for 5 s, whereas the enclosed ZB segment was grown for either 40 or 20~s for the two sets of nanowire growth samples studied in this paper. These samples are referred to as rowth 1 and 2, respectively.

TEM analysis was carried out in a JEOL 3000F setup operated at 300~kV where the nanowires were mechanically broken off the growth substrate before transferring them onto lacy carbon-covered copper grids. A Zeiss Leo Gemini 1560 (in-column situated secondary electron detector) was used for acquisition of SEM data with a typical acceleration voltage of 15~kV and extraction current on the order of 100~$\mu$A. Sample tilts in the range of -5$^{\circ}$ to 20$^{\circ}$ were set to ensure satisfactory conditions for ECCI to distinguish WZ and ZB segments in the nanowires.

Nanowire devices were fabricated by mechanically transferring nanowires from the growth samples to degenerately $n$-doped silicon substrates with a 110-nmthick SiO$_{2}$ film. Such a substrate has predefined gold markers and contact pads, and the back side is covered with gold. The doped Si substrate serves as a global back gate during the electrical measurements. Source and drain contacts consisting of 25-nm Ni and 75-nm Au and were processed by electron beam lithography, followed by lift-off. A 30-s O$_{2}$-plasma etch and an HCl:H$_{2}$O~(1:20) etch for 10~s were carried out prior to metallization to remove resist residue and native oxide, respectively.

\section{Results and Discussion}
\begin{figure}[h]
\centering
\includegraphics[width=0.7\columnwidth]{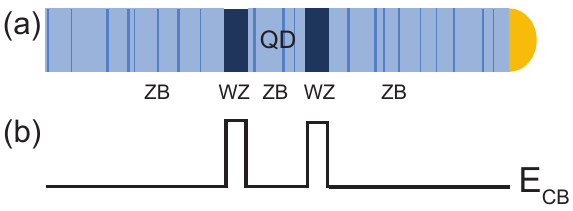}
\caption{(Color online) (a) Schematic of the nanowire crystal structure composition, where the QD is defined by WZ segments (dark blue) in an otherwise ZB (light blue) nanowire. The stripes in the ZB segments represent twinned segments. (b) Sketch of the conduction-band edge $E_{CB}$ of the system, where the WZ segments are represented by simple square potential barriers.}
\label{fig1}
\end{figure}

A schematic representation of the nanowire structure is shown in Fig.~\ref{fig1}(a), together with a sketch of the conduction-band edge $E_{CB}$ alignment based on the simplified assumption that WZ forms a square potential barrier relative to ZB [Fig.~\ref{fig1}(b)]. The WZ segments (dark blue) define a small quantum dot in an otherwise ZB (light blue) nanowire, and the stripes in the ZB segments represent twinned segments, unintentionally incorporated during growth. It was previously reported that such twinned segments do not affect the resistivity in InAs nanowires at 295~K, suggesting that other sources of scattering are limiting transport, such as surface states.\cite{Thelander2011} As will be shown, no discernible transport phenomena that can be attributed to twinned segments are found in this study. 

\begin{figure}[h]
\centering
\includegraphics[width=1\columnwidth]{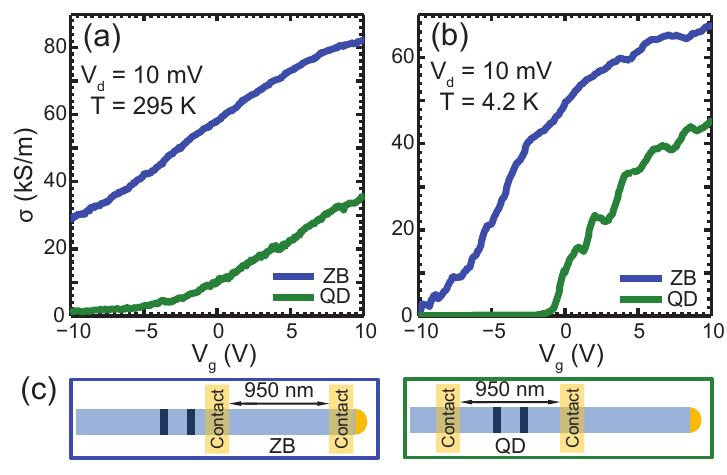}
\caption{(Color online) Conductivity σ as a function of gate voltage $V_{g}$ measured over a QD structure (blue) and over a reference ZB segment (green), at (a) $T= 295$~K and (b) $T = 4.2$~K. The length and diameter of the QD are 45 and 78~nm, respectively, and the thicknesses of the WZ barriers are 22 and 19~nm. The accuracy of the dimensions extracted from SEM analysis is estimated to be $\pm$2 nm. (c) Schematic drawings of the contact configurations for the nanowires measured.}
\label{fig2}
\end{figure}

We find distinct effects of the WZ barriers on conductivity and pinch-off voltage compared to ZB reference samples. Conductivity as a function of gate voltage for a QD device (growth 1) and over a reference ZB segment (growth 1) at 295~K and 4.2~K are displayed in Fig.~\ref{fig2}(a) and (b), respectively. The contact configurations in the two cases are shown in the cartoons in Figs.~\ref{fig2}(c). A clear suppression in conductivity is visible at 295~K for the QD device compared to the reference ZB segment, and at 4.2~K we find a difference in pinch-off voltage of around 10~V. This is in contrast to what was previously reported in Ref.~\onlinecite{Dick2010}, where no dramatic effects on the current levels and pinch-off voltage for QD structures consisting of 4-nm WZ barriers separated by a 130-nm ZB segment were observed. This could be explained by the much thinner WZ barriers used in that study compared to those studied in this work. Measurements on more QD and reference ZB devices at 4.2~K are displayed in Fig.~A.1(a) in the Appendix. Figure~A.1(b) in the Appendix shows SEM images of the reference ZB devices where twinned segments can clearly be seen. This confirms that the WZ segments are responsible for the increased resistance, and that these have a much stronger effect on conductance than twinned segments, which was indirectly shown in Ref.~\onlinecite{Thelander2011}.
\begin{figure*}
\centering
\includegraphics[width=\textwidth]{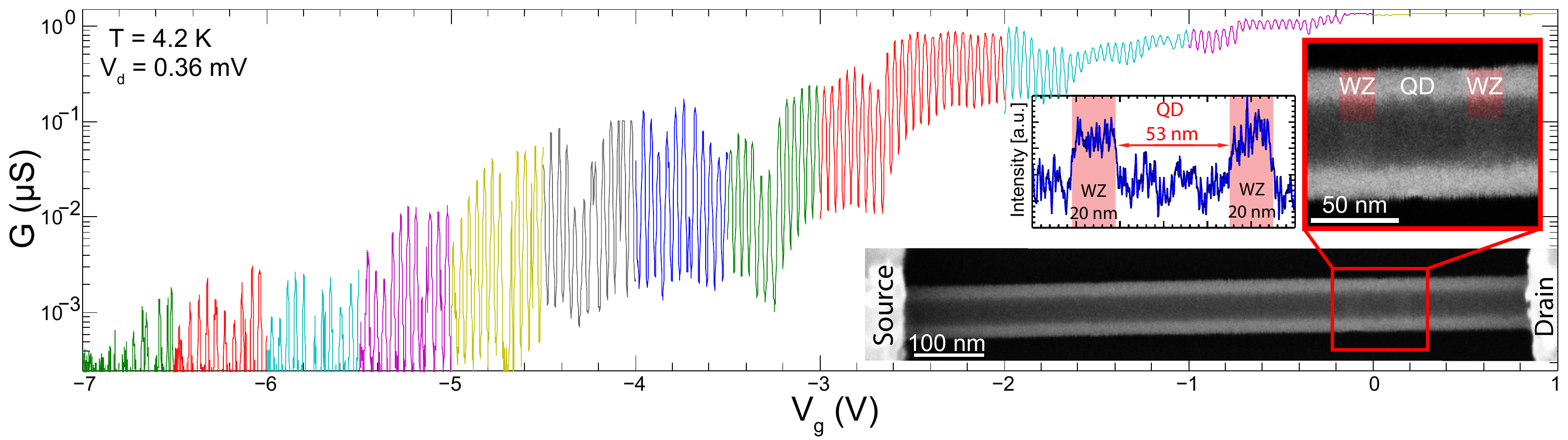}
\caption{(Color online) Composite of sequential conductance measurements as a function of gate voltage, discriminated by color. We note that at the highest current levels the current is limited by the 1-M$\Omega$ series resistance of the current amplifier. Measurements at $V_d =$~10~mV, using a lower series resistance, show a device resistance of only 48 k$\Omega$ at $V_g$~=~0~V, which thus explains the vanishing Coulomb oscillations. Inset: Bottom, top-view SEM image of the corresponding device; top right, a high-resolution SEM of the QD structure where parts of the WZ segments have been colored red to guide the eye; and top left, corresponding intensity profile where the lengths of the QD and barriers are indicated.}
\label{fig3}
\end{figure*}

Next, the role of the WZ segments in low-temperature charge transport is studied. Conductance as a function of gate voltage at a bias of 0.36~mV for a typical QD device (growth 1) at 4.2~K is displayed in Fig.~\ref{fig3}. The graph is a composite of several sequential measurements, following Coulomb oscillations from depletion ($\backsim$-7 V), to a point where the oscillations cease ($\backsim$0 V). Well-defined Coulomb oscillations, observable over a wide range in gate voltage, is a general behavior for the QD devices in this study, which demonstrates that the double WZ segments create significant tunnel barriers in the conduction band. Close to depletion  the Coulomb peaks are somewhat irregularly spaced, which is commonly observed in few-electron QDs.\cite{Kouwenhoven1997} Fourier transform analysis of the Coulomb peaks shows that this initial irregularity is followed by more than a hundred peaks with nearly constant frequency. For more detailed analysis of the gate dependence of the Coulomb oscillations, see Fig.~\ref{figA4} in the Appendix. Figure~\ref{fig3} shows that the conductance overall increases with gate voltage; this is attributed to an increase in both tunnel probability and thermionic emission over the barriers, as well as an onset of cotunneling when the effective barrier height decreases.

The inset in Fig.~\ref{fig3} shows a top-view SEM image of the measured device and a high-resolution SEM image of the QD structure together with an intensity profile. From the high-resolution SEM image, the nanowire and QD dimensions can be extracted using ECCI. The latter indicates a 75-nm-diameter nanowire, with a pair of 20-nm WZ segments enclosing a 53-nm ZB segment. Here we have used an approach that exploits ECCI to deduce the crystal phase directly from SEM images, enabled by careful correlation with information from nanowires studied by TEM only, which we now discuss in more detail.


\begin{figure*}
\centering
\includegraphics[width=\textwidth]{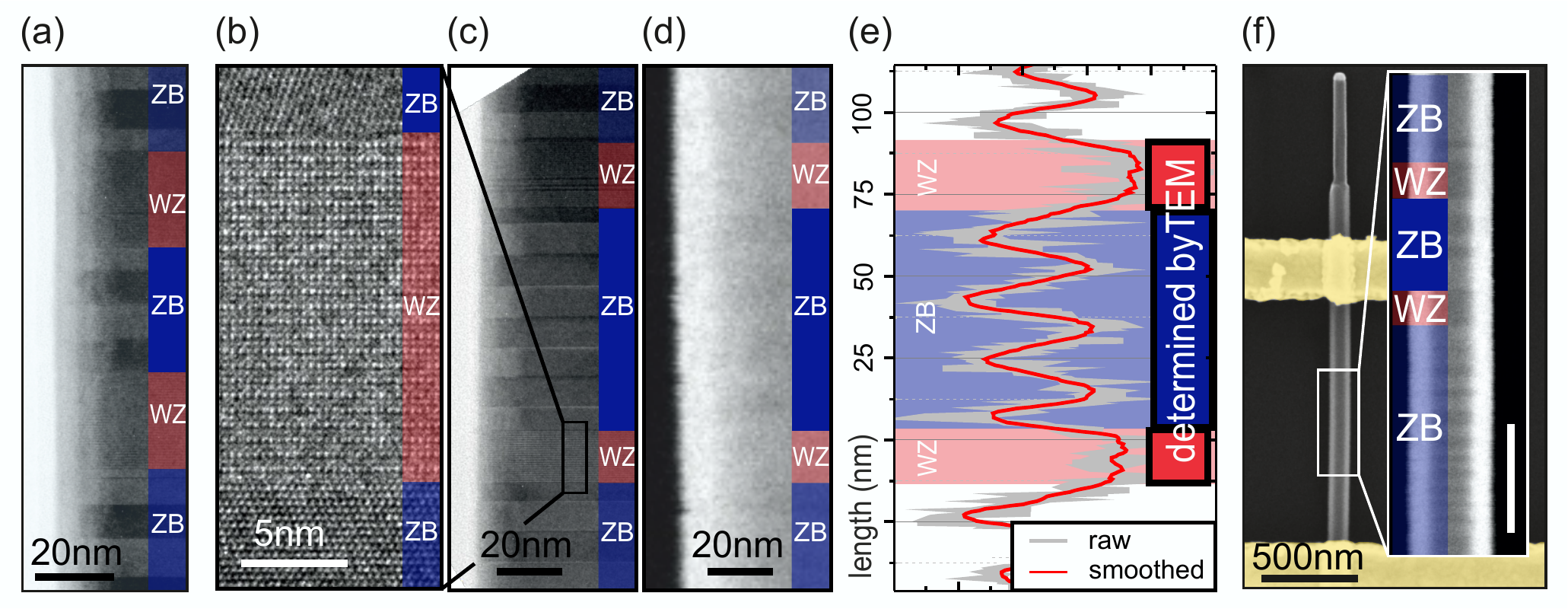}
\caption{(Color online) (a)  TEM image of a nanowire from growth 2, with WZ segment lengths of 25 nm, and a ZB QD length of 32 nm. (c)  TEM and (d)  SEM ECCI data acquired on the same position of the same nanowire from growth 1. In the magnified TEM image in (b) the crystal phase can be distinguished. Dimensions extracted from (c): WZ barrier lengths  16~nm/20~nm and ZB QD length 66~nm. The accuracy of the dimensions extracted from the TEM images is $\pm$1 nm. 
 (e) Extracted intensity profile from the SEM image (d) averaged over 50 lines.  The segment lengths determined by TEM are given in (e) as boxes framed black with a thickness of the frames corresponding to the $\pm$2 nm accuracy of SEM-based segment length extraction. All dimensions extracted by SEM are similar to those extracted by TEM within this error margin.
(f) SEM image of a nanowire device  (growth 1), where crystal phase transitions and twinned ZB segments can be mapped out from the close-up. The scale bar in the close-up is  100~nm. The following dimensions were extracted: WZ segment lengths of 35~nm and a ZB QD length of 81~nm. For electrical measurements on this device, see Figs.~\ref{fig5}(a) and (e). }
\label{fig4}
\end{figure*}

Figures~\ref{fig4}(a) and (c) display high-resolution TEM images of nanowires from growth 2 (shorter QD) and growth 1 (longer QD), respectively. Blue (ZB) and red (WZ) shadings are used to indicate the crystal phase of the nanowire; the darker blue indicates the QD. In the close-up in Fig.~\ref{fig4}(b), the atomically sharp transition between the crystal phases is clearly visible. Nanowires from both growth samples exhibit a ZB phase with rotational twinning perpendicular to the [$\bar{1} \bar{1} \bar{1}$] growth axis of the nanowire, which can be seen as bright and dark contrast areas in the TEM images. 

Figure~\ref{fig4}(d) shows an SEM image of the same nanowire imaged by TEM in Fig.~\ref{fig4}(c). Aligning the electron beam with crystal planes in the nanowire allows the crystal phases to be distinguished and the QD dimensions to be extracted, thus making it possible to directly correlate electrical properties with structural properties.
Since the nanowire in Fig.~\ref{fig4}(d) is placed on a TEM grid for imaging, the quality of the SEM image in Fig.~\ref{fig4}(d) is lower compared to images of nanowire devices on silicon substrate [such as Fig.~\ref{fig4}(f)] due to vibrational instability. However, the dimensions of the segments in Fig.~\ref{fig4}(d) can be extracted from the intensity profile in Fig.~\ref{fig4}(e). For all dimensions extracted from SEM analysis in this work, we estimate an accuracy of $\pm$2 nm.

Depending on the alignment of the nanowire with respect to the electron beam, the contrast between WZ and ZB can vary. Furthermore, the conditions at which twinned ZB segments can be distinguished do not necessarily overlap with conditions where it is possible to distinguish between ZB and WZ; for more details see the Appendix.

We note that the use of randomly distributed aerosol seed particles in the growth results in a substantial spread in the QD dimensions. However, this spread is not a limitation in this study since the dimensions can be extracted from SEM analysis after the electrical measurements. There are two major factors that affect the QD dimensions and segment lengths: First, there is a direct correlation between seed particle size and nanowire diameter, such that a spread in gold-particle diameter will give a corresponding spread in nanowire growth rate and quantum dot dimensions.\cite{Froberg2007} Second, nanowire growth rate is also affected by local variations in particle density on the growth substrate.\cite{Froberg2007,Borgstrom2007}


\begin{figure*}
\centering
\includegraphics[width=\textwidth]{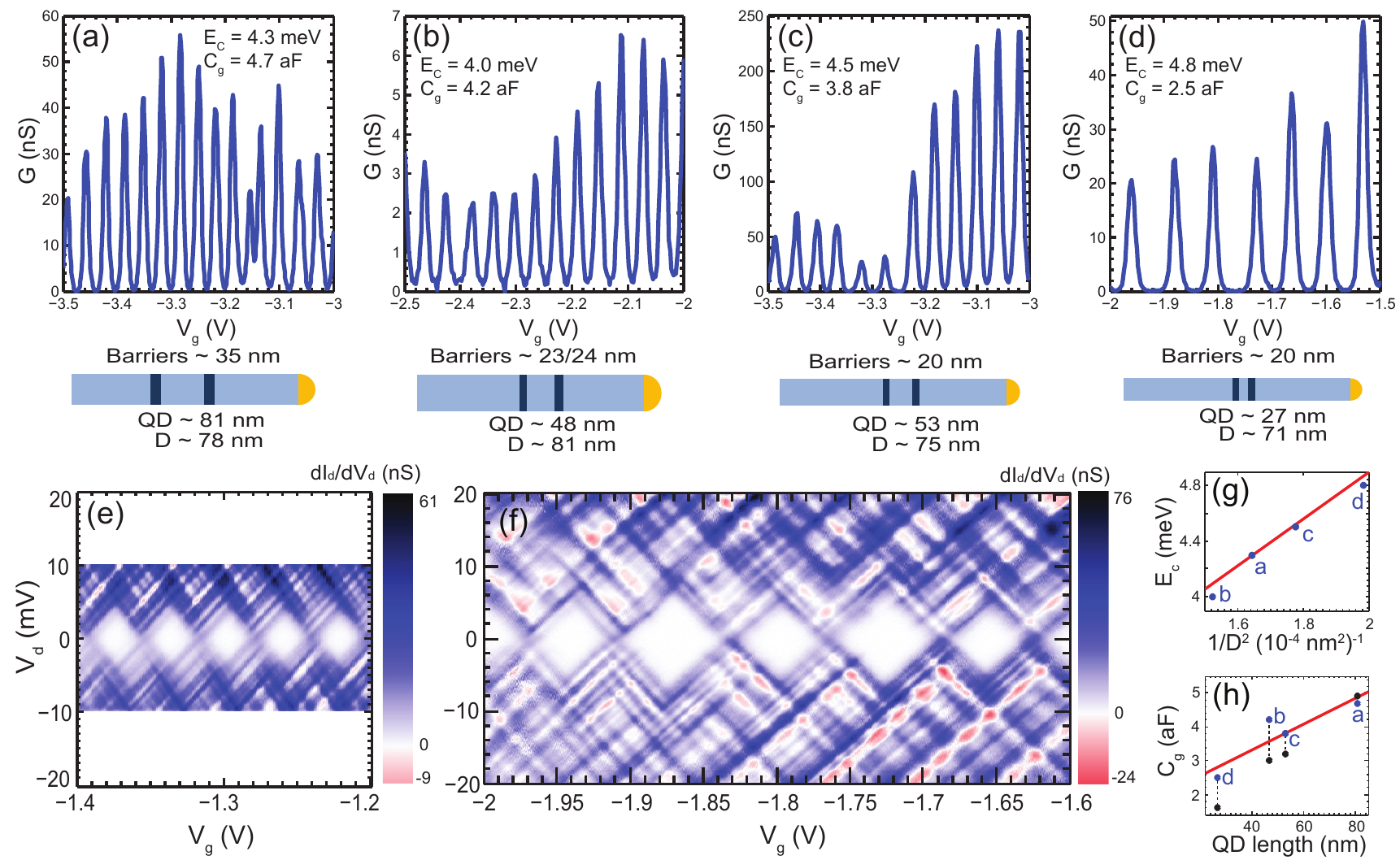}
\caption{(Color online) Coulomb oscillations in conductance as a function of gate voltage for QDs with different lengths, (a)~81~nm,  (b)~48~nm, (c)~53~nm (same device as in Fig.~\ref{fig3}), and (d) 27~nm, and cartoons (not to scale) of corresponding QD structures. In the cartoons, "Barriers" denotes the width of the WZ segments (dark blue), "QD" denotes the QD length and "D" denotes the nanowire diameter. All measurements were performed at $T = 4.2$~K with a bias of $V_{d} = 0.36$~mV. (e), (f) Charge stability diagrams of same devices as (b) and (d), respectively, at $T = 4.2$~K. Positive conductance lines running perpendicular to the diamond edges are visible in both stability diagrams, and correspond to the onset of sequential tunnel transport via excited states. In addition to these lines, negative conductance lines running parallel to the excited states are visible, which we attribute to fluctuations in tunnel probability between conducting modes in the leads and states in the QD as mentioned in the text. (g)  Extracted  charging  energy vs. the inverse of the nanowire diameter square. (h) Gate capacitance (blue, experimental; black, analytical)  vs length of the QD for the four devices, respectively. The red lines in (g) and (h) are  linear fits to the experimental data points.  }
\label{fig5}
\end{figure*}
Next, the electrical properties of several QDs at \textit{T}~=~$4.2$~K are correlated to the QD-dimensions extracted from SEM images. Figures~\ref{fig5}(a)~-~(d) show measurements of conductance as a function of gate voltage recorded for four QD devices with different dimensions, as well as cartoons of the corresponding QD structures. Figure~\ref{fig5}(a)-(c) represent nanowires grown with a longer growth time (40~s) for the ZB QD segment, whereas Fig.~\ref{fig5}(d) represents a nanowire grown with a shorter growth time (20~s) for the QD segment. Figures~\ref{fig5}(e) and (f) show charge stability diagrams recorded for the same devices as displayed in Fig.~\ref{fig5}(b) and (d), respectively.

Regular Coulomb oscillations are clearly visible for all QD devices in Figs.~\ref{fig5}(a)-(d), where the oscillations continue over gate voltage ranges of 5-7~V, showing that the WZ segments introduce a significant conduction band offset. As expected for QD systems in the quantum Coulomb blockade regime ($kT \ll$ single particle energy $E_{\Delta}$, and charging energy $E_{C}$), the amplitude of the Coulomb peaks is gate-voltage dependent. This dependence is due to different tunnel coupling strengths between conducting modes in the leads and states on the dot, in combination with an energy-dependent density of states in the leads.

In the case of the three QDs in Figs.~\ref{fig5}(a)-(c), the Coulomb oscillation frequency is nearly constant, indicating that $E_{C}$ is much larger than $E_{\Delta}$, and is the dominating term in the addition energy ($E_{Add} = E_{C} + E_{\Delta}$). In Fig.~\ref{fig5}(e), this is evident from the equal size of the Coulomb diamonds. From charge stability diagrams, charging energies ($E_{C}= e\Delta V_{d}$) of 4.3, 4.0, and 4.5~meV and nearly constant gate capacitances ($C_{g}= \frac{e}{\Delta V_{g}}$ ) of approximately 4.8, 4.2, and 3.8~aF are extracted for the three devices, respectively. Here, $\Delta V_{d}$ and $\Delta V_{g}$ are the maximum height (measured from $V_{d}=0$) and the total width (measured at $V_{d}=0$) of the Coulomb blockade diamond, respectively.

When the length of the QD is reduced to 27~nm, the contribution of $E_{\Delta}$ can be resolved as irregular Coulomb peak spacings with an odd-even behavior [Fig.~\ref{fig5}(d)], which is also clearly visible in the charge stability diagram [Fig.~\ref{fig5}(f)]. In this case, a charging energy of approximately 4.8~meV, and a reduced gate capacitance of 2.5~aF, as well as a constant  $E_{\Delta}$  of 1.4~meV,  are extracted. In the case of a stability diagram recorded at more negative gate voltages (-3~to~-2.6~V), the extracted  $E_{\Delta}$ increases to 2-4~meV. 

\begin{table*}
 \caption{Extracted dimensions, changes in the number of electrons ($\Delta$ electrons) and electron concentration $n$ in the QD at the point where Coulomb oscillations cease, and calculated barrier heights ($E_{F}-E_{CB}$) for three QD devices. All three devices consisted of nanowires from growth 1.} \label{tab1}
\begin{ruledtabular}
\begin{tabular}{c c c c c c c c}
 \rule{0pt}{3ex}Device  & Diameter  & QD length  & WZ barrier lengths  & $\Delta$  electrons  & $n$  & Barrier height  \\ 
&(nm)&(nm) & (nm) & &(cm$^{-3}$) &(meV)\\[2pt]

 \hline
\rule{0pt}{3ex}   I & 83 & 61 & 33/37 &  217 & $6.6\times10^{17} $& $>$ 95 \\ 

\rule{0pt}{3ex}II [Fig. \ref{fig3}, Fig. \ref{fig5}(c)] & 75 & 53 & 20 & 159 & $6.8\times10^{17} $ & $>$ 97 \\ 

\rule{0pt}{3ex}III [Fig. \ref{fig2}] & 78 & 45 & 22/19 &  139 & $6.5\times10^{17} $ & $>$ 94 \\ 
\end{tabular} 
\end{ruledtabular}
\end{table*}

In order to correlate the dimensions of the QD devices with their electrical properties, the charging energy vs. $1/D^2$, where $D$ is the nanowire diameter, are shown in Fig.~\ref{fig5}(g). The charging energy  can be written as $E_{C} = e^{2}/C$, where $C$ is the total capacitance of the  QD. In Fig.~\ref{fig5}(g), $E_{C}$ shows a linear dependence with $1/D^2$, which is expected for disc-shaped QDs where the source and drain capacitances dominate. Figure~\ref{fig5}(h) shows the experimentally and analytically extracted gate capacitances (blue and black)  vs length. Here, an analytic model for estimating $C_g$ of a metallic, infinitely long cylinder on a plane has been used: \cite{Khanal2006}
\begin{equation}
C_g = \frac{2\pi\epsilon_0\epsilon_rL}{cosh^{-1}(\frac{D}{2}+h/\frac{D}{2})},
\end{equation}
where  $L$ is the nanowire length (here replaced with QD length), $h$ is the gate oxide thickness (110~nm) and $\epsilon_r$ is  the relative dielectric constant of the oxide.  For the latter we use a reduced $\epsilon_r$ of  2.2 to take into account that the nanowire is not embedded in SiO$_2$. \cite{Wunnicke2006} The trend in Fig.~\ref{fig5}(h) shows a scaling of $C_{g}$ with the length of the QD segment, in good agreement with the analytical expression, clearly indicating that the WZ segments act as tunnel barriers and effectively define a QD.


We now extract a lower limit for the WZ conduction barrier offset relative to ZB by estimating the electron population and concentration on the QD from the number of Coulomb oscillations observed. For sufficiently high $V_g$, the WZ segments become too transparent (R~$<$~26~k$\Omega$/barrier) to localize electrons on the embedded ZB segment. Here, the position of the Fermi level $E_F$ relative to the ZB conduction-band edge $E_{CB}$ provides a lower estimate of the conduction-band offset. In the opposite limit, at the onset of Coulomb oscillations, we first observe irregular Coulomb blockade peaks. This indicates that some parts of the nanowire (not necessarily the QD) are close to depletion. Here we introduce the other lower bound for the conduction-band offset, by assuming a Fermi level position very close to the ZB conduction-band edge. The electron density $n$ on a dot is now calculated from the number of observed Coulomb oscillations (between the limits discussed above) divided by the QD volume extracted from an SEM image, using a circular cross section of the QD. By evaluating the following expression for the electron
density, where the nonparabolicity of the band is compensated for, ($E_{F}-E_{CB}$) can be extracted:\cite{Ariel-Altschul1992,Lind2010} 

\begin{equation}
n = \frac{2N_{C}}{\sqrt{\pi}}\int_{0}^{\infty}\frac{\sqrt{\varepsilon(1+\alpha\varepsilon)}(1+2\alpha\varepsilon)}{1+exp(\varepsilon-\phi)}d\varepsilon.
\label{eq1}
\end{equation}

Here, $N_{C}$ is the effective density of states in the conduction band,  $\varepsilon=(E-E_{CB})/kT$ is the normalized electron kinetic energy, $\phi=(E_{F}-E_{CB})/kT$ is the normalized Fermi energy, and $\alpha=(1-\frac{m_{e}}{m_{0}})^{2}/\varepsilon _{g}$ is the nonparabolicity factor, containing the effective mass $m_{e}$ and the normalized band gap  $\varepsilon _{g}= E_{g}/kT$. The system is assumed to be in the extreme degeneracy or low-temperature regime $E_{F}-E_{CB}\gg kT = 0.36$~meV) where a step-function approximation of the Fermi function in the evaluation of Eq.~(\ref{eq1}) can be used. In this estimation, we have assumed that the effect of the gate on the conduction-band edge is constant in all ZB segments, and quantum confinement has been neglected.

Three devices have been analyzed, which resulted in lower estimations of the barrier heights of 95, 97, and 94~meV, respectively; details of the devices are stated in Table~\ref{tab1}.  For InAs $\lbrace110\rbrace$-type surfaces with a thin oxide, the Fermi level is reported to be pinned approximately 100~meV above the conduction-band edge.\cite{Baier1986,Tsui1970} This is consistent with the Fermi level position at $V_{g} = 0$~V found here, which coincidentally also is the point at which the Coulomb blockade due to the WZ segments is lifted in most samples studied.

\section{Summary}
In summary, we demonstrate controlled formation of QDs in nanowires through polytypic design. The QDs are formed by introducing pairs of thin, closely spaced segments of WZ in otherwise ZB InAs nanowires. Coulomb oscillations with reproducible periodicity are observed over a wide gate-voltage span, indicating a significant conduction-band offset. A method is developed to extract QD dimensions and barrier widths directly from SEM images, which is used in the analysis of the electrical measurement data. With this information, it is possible to directly correlate electrical properties of the QDs, such as gate capacitance and charging energy, with the dimensions of the crystal phase segments. Confinement effects can be clearly resolved for the shortest QD lengths where $kT~\ll~E_{\Delta}$ at 4.2~K. Based on estimations of the carrier concentrations in longer QDs by counting Coulomb oscillations, we extract a lower bound for the ZB-WZ conduction-band offset of  95~meV. In view of recent findings that WZ nanowire surfaces can effectively inhibit radial overgrowth,\cite{Namazi2015} we finally conclude that polytype control may offer a path to realize more complex systems, such as epitaxially designed core-shell QDs.  

\begin{acknowledgments}
This work was carried out with financial support from NanoLund, the Swedish Research Council (VR), the Swedish Foundation for Strategic Research (SSF), and the Knut and Alice Wallenberg Foundation (KAW).
\end{acknowledgments}

\appendix*
\setcounter{figure}{0} \renewcommand{\thefigure}{A.\arabic{figure}}
\section{}
\begin{figure*}
\centering
\includegraphics[width=0.8\textwidth]{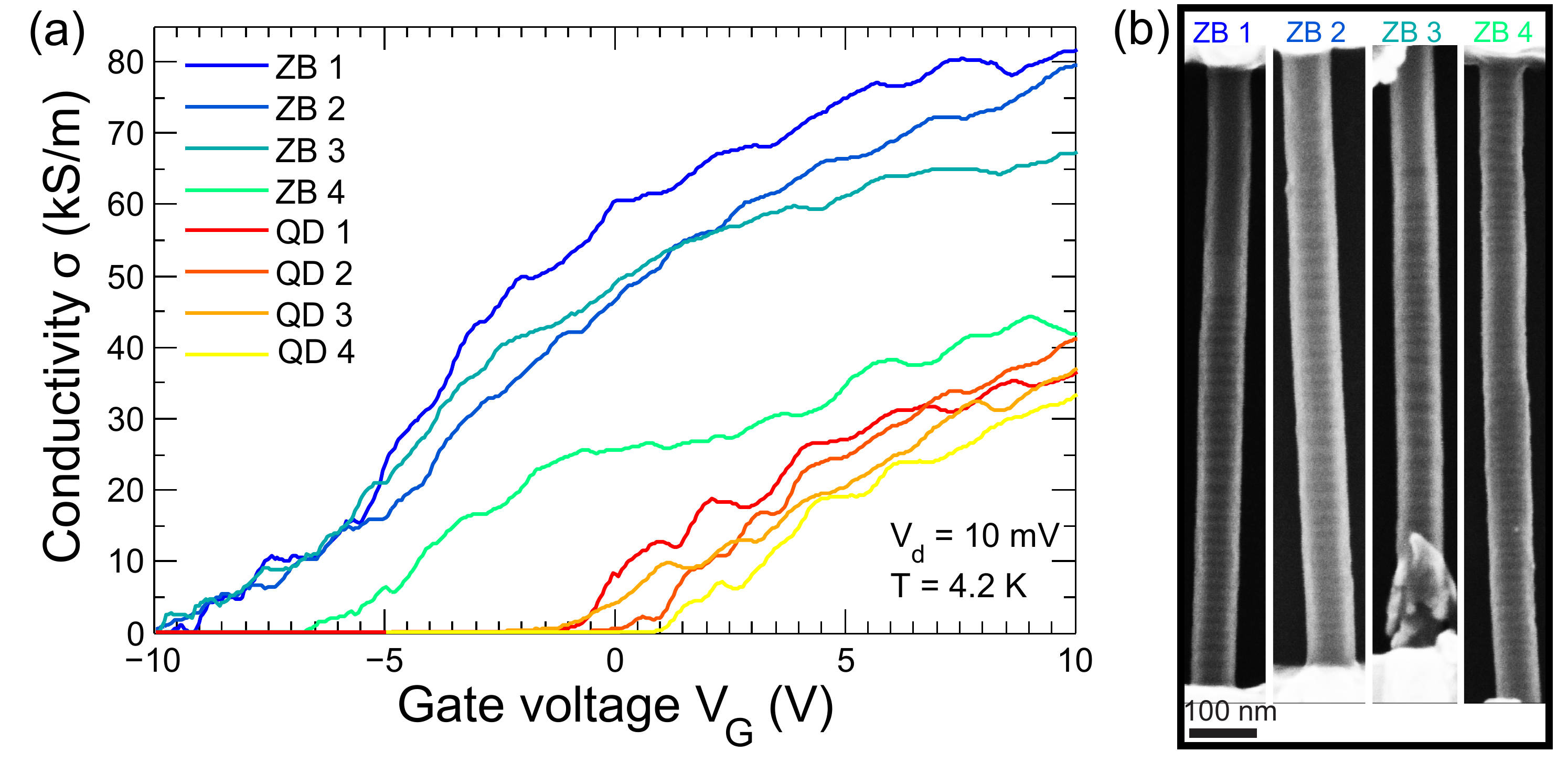}
\caption{(Color online) (a) Conductivity as a function of gate voltage for four reference ZB segments (blue-green color scale) and for four QD devices (red-yellow color scale). (b) SEM images of the ZB reference devices optimized to observe the twinned segments.}
\label{figA1}
\end{figure*}

For statistics on the impact of the QD structure on the conductivity, see Fig. \ref{figA1}(a), where conductivity is plotted as a function of gate voltage for four reference ZB devices and four QD devices. Figure \ref{figA1}(b) shows SEM images of the reference ZB devices with visible contrast originating from twinned segments, strongly indicating the weak impact of the twinned segments on the current level compared to the WZ segments in the QD devices.

\begin{figure*}
\centering
\includegraphics[width=0.7\textwidth]{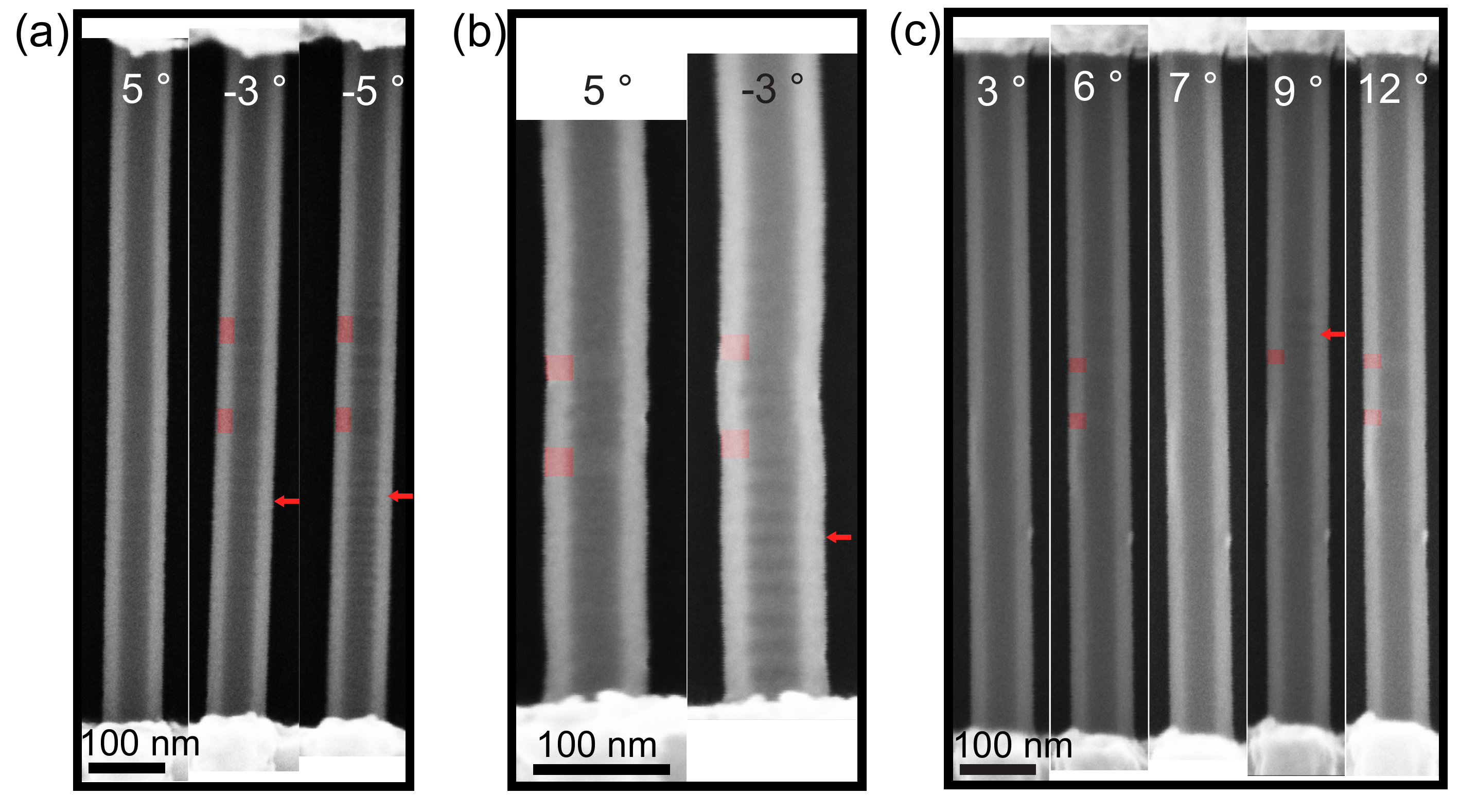}
\caption{(Color online) (a)-(c) SEM images of three different QD devices at different tilt angles of the SEM stage. When visible, the WZ segments are colored red, and tilt angles where twinned segments in ZB are visible are indicated by arrows. }
\label{figA2}
\end{figure*}

Figures~\ref{figA2}(a)-(c) show SEM images with different tilt angles of the sample holder for three different QD devices. At optimum tilt angle, the WZ segments can give rise to dark or bright contrast depending on the device. In Fig.~\ref{figA2}(a) both the WZ segments (red) and the twinned segments in ZB (arrow) appear when adjusting the tilt angle. In contrast, in Fig.~\ref{figA2}(b) no twinned segments are visible at the optimal tilt angle ($5^{\circ}$) for observing the WZ segments. The twinned segments instead appear at a different angle ($-3^{\circ}$). In the case of Fig.~\ref{figA2}(c) the WZ segments are visible at $6^{\circ}$ and $12^{\circ}$, whereas twinned segments appear at $9^{\circ}$. Here it should be mentioned  that the extracted dimensions are very sensitive to drift during imaging. In order to limit the error additional fast-scan-speed images have been used when extraction dimensions.

\begin{figure*}
\centering
\includegraphics[width=0.6\textwidth]{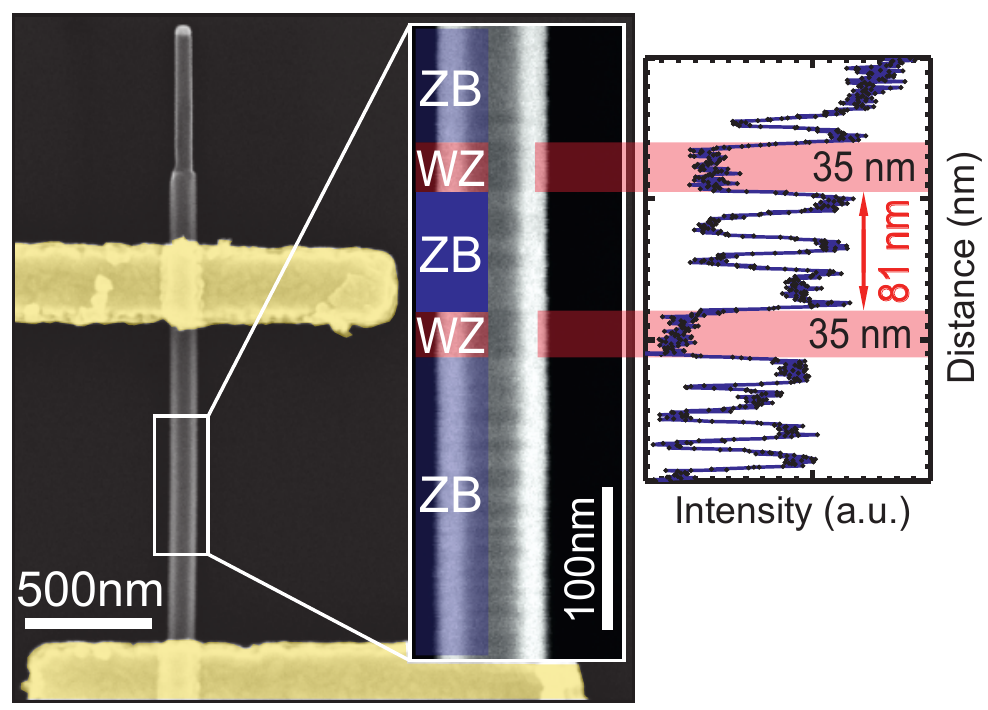}
\caption{(Color online) (a) SEM images from Fig.~\ref{fig4}(f) in the main text with (b) corresponding intensity profile. }
\label{figA3}
\end{figure*}

Figure~\ref{figA3}(a) shows the same SEM images as Fig.~\ref{fig4}(f) in the main text with an addition of an intensity profile Fig.~\ref{figA3}(b) along the growth direction of the nanowire.

\begin{figure*}
\centering
\includegraphics[width=0.4\textwidth]{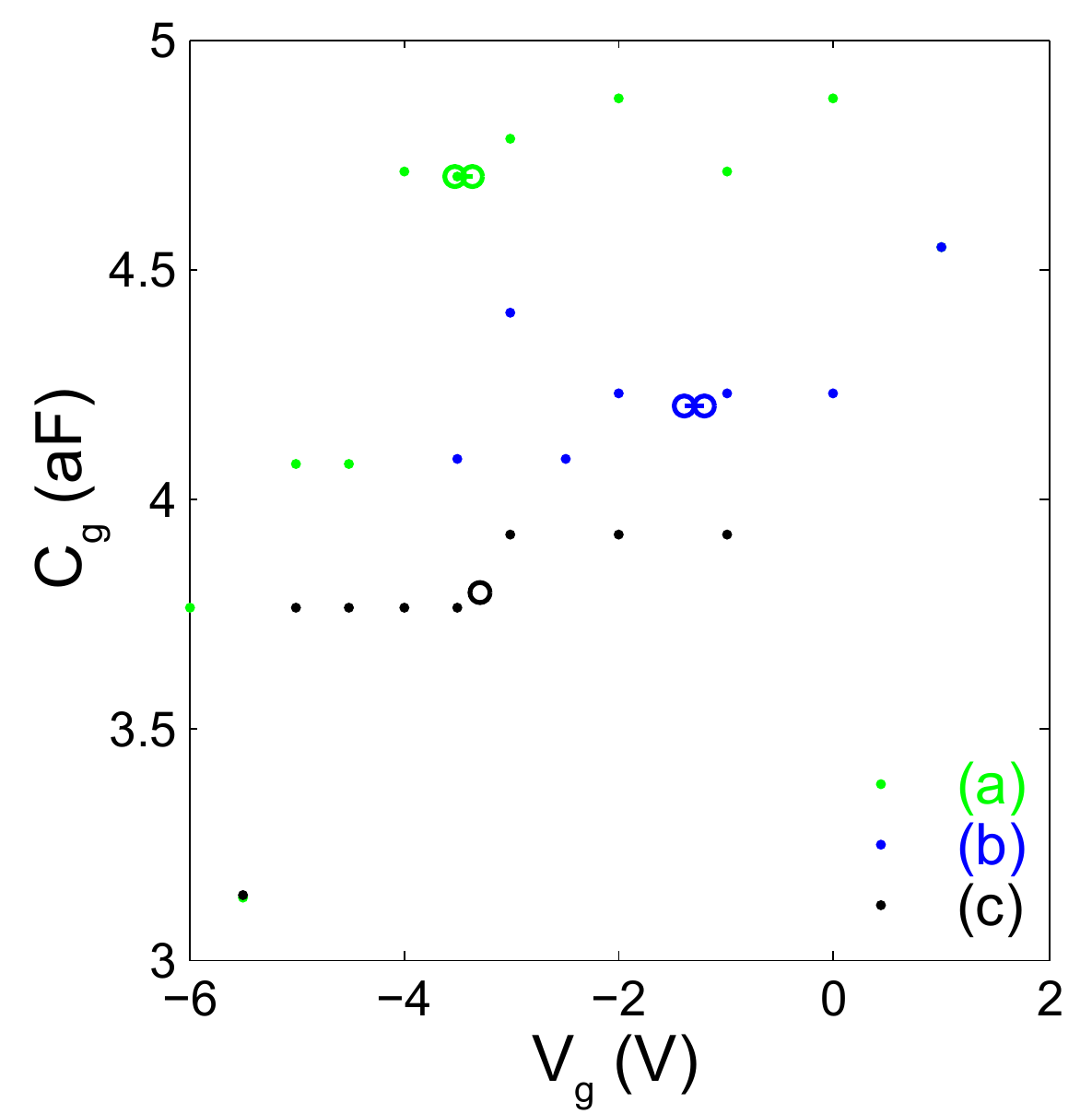}
\caption{(Color online) Gate capacitance as a function of gate voltage. Results from fast Fourier transform (FFT) analysis (dots) where the gate sweep has been divided into subintervals when performing FFT; the point in the graph indicates the start of one such interval. The circles indicate from which gate-voltage range the $C_g$ displayed in Fig.~\ref{fig5} are  extracted. }
\label{figA4}
\end{figure*}

A detailed analysis of the Coulomb oscillation period as a function of gate voltage is presented in Fig.~\ref{figA4}. Going from positive to negative $V_g$ we observe a small drop (10-20 $\%$) over the full gate-voltage interval. A reduced gate capacitance at negative gate voltages may, in part, be related to a change in barrier width or  a small shift of the quantum dot center away from the gate.


\bibliography{My_Collection}

\end{document}